\documentclass[a4paper]{article}

\usepackage[switch,columnwise]{lineno}
\usepackage{colortbl}
\usepackage{graphicx}
\usepackage{geometry}
\usepackage{algpseudocode}
\usepackage{pifont}
\usepackage[percent]{overpic}
\usepackage{amsfonts}
\usepackage{multirow}
\usepackage{url}
\usepackage{tabularx}

\usepackage{collcell}
\usepackage{hhline}
\usepackage{pgf}

\def\colorModel{hsb} 

\newcommand\ColCell[1]{
	\pgfmathparse{#1<50?1:0}  
	\ifnum\pgfmathresult=1\relax\color{white}\fi
	\pgfmathsetmacro\compA{0}      
	\pgfmathsetmacro\compB{0} 
	\pgfmathsetmacro\compC{#1/100}      
	
	\edef\x{\noexpand\centering\noexpand\cellcolor[\colorModel]{\compA,\compB,\compC}}\x #1
} 
\newcolumntype{E}{>{\collectcell\ColCell}m{1cm}<{\endcollectcell}}  

\renewcommand{\vec}[1]{\boldsymbol{\mathbf{#1}}}

\usepackage[english]{babel}
\usepackage[utf8]{inputenc}
\usepackage{amsmath}
\usepackage{amsthm}
\usepackage{colortbl}
\usepackage{graphicx}
\usepackage{color}
\usepackage{algorithm}
\usepackage{pifont}
\usepackage[english]{babel}
\usepackage[utf8]{inputenc}
\usepackage{url}
\usepackage{amsmath}
\usepackage{tabularx}
\usepackage{collcell}
\usepackage{hhline}
\usepackage{pgf}
\usepackage{graphicx}
\usepackage[colorinlistoftodos]{todonotes}
\usepackage{geometry}
\usepackage{subfigure}
\usepackage{caption}
\usepackage{algorithm}
\usepackage{algpseudocode}
\usepackage{pifont}
\usepackage{multirow}
\usepackage{amsfonts}
\usepackage{lineno}
\usepackage[percent]{overpic}
\usepackage{todonotes} 
\usepackage{soul} 
\definecolor{myc}{HTML}{3A08C8}
\usepackage[toc,page]{appendix}

\renewcommand{\vec}[1]{\boldsymbol{\mathbf{#1}}}

\title{Automatic scoring of apnea and hypopnea events using blood oxygen saturation signals}

\author{R.E. Rolon$^{a,d}$, I.E. Gareis, L.D. Larrateguy$^b$, L.E . Di Persia$^a$, R.D. Spies$^c$ and H.L. Rufiner$^a$}

\date{\today}

\begin{document}

\maketitle

\begin{center}
$^a$ Instituto de Investigaci\'{o}n en Se\~{n}ales, Sistemas e Inteligencia Computacional, sinc($i$)-UNL-CONICET, Santa Fe, Argentina
	
$^b$ Centro de Medicina Respiratoria de Paran\'{a}, Entre R\'{i}os, Argentina

$^c$ Instituto de Matem\'{a}tica Aplicada del Litoral, IMAL-UNL-CONICET, Santa Fe, Argentina
	
$^d$ Facultad Regional Paran\'{a}, Universidad Tecnol\'{o}gica Nacional, Entre R\'{i}os, Argentina
	
\end{center}

\begin{abstract}
The obstructive sleep apnea-hypopnea (OSAH) syndrome is a very common and frequently undiagnosed sleep disorder. It is characterized by repeated events of partial (hypopnea) or total (apnea) obstruction of the upper airway while sleeping. This study makes use of a previously developed method called DAS-KSVD for multiclass structured dictionary learning to automatically detect individual events of apnea and hypopnea using only blood oxygen saturation signals. The method uses a combined discriminant measure which is capable of efficiently quantifying the degree of discriminability of each one of the atoms in a dictionary. DAS-KSVD was applied to detect and classify apnea and hypopnea events from signals obtained from the Sleep Heart Health Study database. For moderate to severe OSAH screening, a receiver operating characteristic curve analysis of the results shows an area under the curve of 0.957 and diagnostic sensitivity and specificity of 87.56\% and 88.32\%, respectively. These results represent improvements as compared to most state-of-the-art procedures. Hence, the method could be used for screening OSAH syndrome more reliably and conveniently, using only a pulse oximeter.
\end{abstract}
Keywords:
Pulse oximetry, apnea-hypopnea events, sleep disorders screening, structured dictionary learning, discriminant measures, multiclass classification problems.


\section{Introduction}
\label{introduction}

Pulse oximetry, being a cheap and non-invasive technique, has become a promising supporting tool for the diagnosis of sleep disorders \cite{yadollahi_sleep_2010,schlotthauer_screening_2014,rolon_discriminative_2017}. Sleep disorders comprise several types of medical conditions. The most common one of them is the Obstructive Sleep Apnea-Hypopnea (OSAH) syndrome, which is caused by frequent breathing pauses due to partial (hypopnea) or total (apnea) blockage of the upper airway during sleeping, which lead to several physiological changes such as blood oxygen desaturation \cite{ Hedner635, berry2012rules}. To establish the severity of this pathology, the apnea-hypopnea index (AHI) is commonly used. This index is defined as the number of apnea-hypopnea events per hour of sleep or record according to whether it refers to a complete study or a simplified one, respectively (more on this later). Most screening methods do not discriminate between apnea and hypopnea events since it is not strictly required for computing the AHI index \cite{schlotthauer_screening_2014}. However, recognition of single apnea and hypopnea events provides additional information regarding the severity of OSAH syndrome that could be important for clinical and decision-making purposes \cite{khadadah2017does}. Nevertheless, automatically distinguishing and identifying those two respiratory events is a challenging task, specially when the number of available signals is low. 

Achieving a good AHI estimation using recordings of just a few signals is a difficult problem that requires of precise ad-hoc evaluation tools for the clinical screening of OSAH syndrome \cite{gamaldo2018evaluation}. In the past decade much interest in the development of portable devices using at most two sensors for OSAH screening has been observed (e.g. \cite{uddin2018classification,del2018oximetry,mendoncca2018review,terrill2019review}). In particular, the authors in \cite{del2018oximetry} present a detailed review of existing methods that use only pulse oximetry signals for automatically classifying patients having OSAH syndrome. It is important to highlight however that all methods mentioned in that review address only the detection of the pathology and do not recognize nor classify small segments of oximetry signals as normal breathing, apnea or hypopnea events. In that way, up to our knowledge, the problem of individually classifying abnormal respiratory events using only $\textrm{SaO}_2$ signals in a multi-class scenario has never been explored before. Therefore, properly identifying hypopneas which were not detected by other approaches may add value in the diagnosis and tratement of the patiens.

There are methods for binary classification (existence or nonexistence of abnormal respiratory events) of $\textrm{SaO}_2$ signals from which the AHI index can be estimated \cite{schlotthauer_screening_2014,rolon_discriminative_2017,chiner_nocturnal_1999,vazquez_automated_2000}. In particular, the articles \cite{chiner_nocturnal_1999} and \cite{vazquez_automated_2000} make use of the so called Oxygen Desaturation Index (ODI) defined as the number of times that the $\textrm{SaO}_2$ signal falls below a prescribed  percentage of signal saturation regarding a baseline level per hour of study. It is timely to point out however that although the concept of ``baseline level'' is somewhat intuitive, there is yet no consensus about its formal definition, and different authors have adopted different ones \cite{chiner_nocturnal_1999,vazquez_automated_2000}. In \cite{chiner_nocturnal_1999}, for instance, the baseline level was defined as the desaturation mean of the previous minute, while a completely different approach was followed in \cite{vazquez_automated_2000} where it was computed using a moving time average. In \cite{schlotthauer_screening_2014}, the authors present a method for detecting blood oxygen desaturations using specific waves (or modes) coming from empirical mode decompositions of $\textrm{SaO}_2$ signals. In that work, the desaturations are identified by making use of a few thresholds and a set of simple rules which lead to the detection of the sleep apnea-hypopnea syndrome. Finally, in \cite{rolon_discriminative_2017}, we introduced a different approach based on sparse representations of $\textrm{SaO}_2$ signals. In that work, the AHI index is directly estimated without computing the ODI index, as the average of the number of abnormal respiratory events per hour of study.

To tackle the problem of individually identifying and distinguishing between apnea and hypopnea events using only $\textrm{SaO}_2$ signals we make use of a previously developed method \cite{rolon_a_multiclass_2018}. For that, segments of training signals are used to learn a discriminant dictionary. Also, at the dictionary learning stage, a multi-class multi-objective information measure is used for quantifying the discriminability of each atom in the dictionary. Finally, sparse representations of the data in terms of the dictionary are computed and then used as input of a classifier (see Section \ref{sec: neural network}).

The organization of this paper is as follows: In Section \ref{sec: sleep apnea}, a brief description about abnormal respiratory events during sleep is presented. Dictionary learning methods for sparse representation are introduced in Section \ref{sec: dict learning for sr}. Section \ref{sec: exp setup} contains details on all designed experiments. Results and discussions are introduced in Section \ref{sec: results and discussions} while conclusions are finally presented in Section \ref{sec: conclusions}.
\section{Sleep apnea}\label{sec: sleep apnea}
It is well known that getting enough sleep is extremely important for maintaining both mental and physical health. However, good sleeping very often becomes affected by the presence of sleep-related breathing disorders. Poor sleep quality causes excessive daytime sleepiness affecting the productivity and efficiency of people, including their ability to think clearly, react quickly and memorize efficiently, triggering bad decisions and highly increasing the risk of having domestic, work and traffic accidents \cite{medic2017short}.

Polysomnography (PSG) is the reference study for diagnosing OSAH syndrome. This study requires of specially conditioned sleep units as well as the simultaneous recording of several biomedical signals. However the accessibility to PSG is very limited mainly because PSG units are not commonly available and because the studies are both lengthy and costly, making the process of obtaining good quality signals extremely complicated. In addition, a PSG study requires the attention of specialized technicians to ensure continuous time visualization and recording of all the signals being acquired. A complete PSG study consists of the simultaneous measuring of a minimum of seven physiological signals such as electroencephalography (EEG), electrooculography (EOG), electromiography (EMG), electrocardiography (ECG), airflow and $\textrm{SaO}_2$. It is important to point out however that the continuous acquisition of these signals highly affects the quality of sleep, making it even more difficult to achieve an accurate diagnosis. Because all those difficulties, new screening approaches are always been developed. An ideal screening method can be considered as one that, on one hand leads to precise results, and on the other hand it uses as few signals as possible without degrading the quality of sleep \cite{gamaldo2018evaluation}.

For the reasons described above, portable systems for assessing OSAH syndrome, that can be used outside sleep units, have been developed. In this sense other evaluation procedures exist, such as home PSG, home Respiratory Poligraphy (RP) and other simplified procedures, to name a few. Although home PSG has the advantage of not requiring of any trained personnel, it still needs the acquisition of at least seven respiratory and sleep signals, just like a standard PSG. On the other hand, home RP studies allow for the evaluation of cardiorespiratory variables without taking into account EEG, EOG and EMG signals and therefore they are unable to detect wakefulness and to determine sleep stages \cite{garcia-diaz_respiratory_2007}. Hence, even though home RP is simpler than both standard PSG and home PSG, it still needs the continuous measurement of several physiological signals, whose acquisition affects sleep quality. Finally, simplified procedures make use of only one or two cardiorespiratory variables, such as airflow, respiratory movements, heart rate, tracheal sound and $\textrm{SaO}_2$. In particular, the $\textrm{SaO}_2$ signal has become a reasonable alternative for OSAH syndrome screening and it is the one that will be used in this article  \cite{yadollahi_sleep_2010,schlotthauer_screening_2014,rolon_discriminative_2017}.

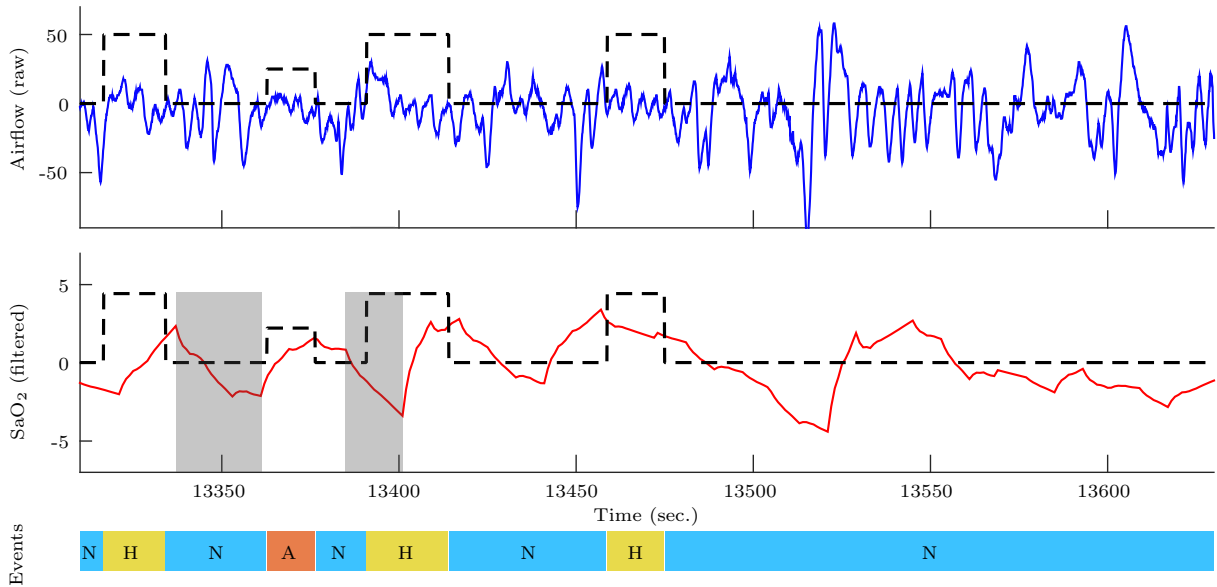
\begin{figure*}[tbp]
	\centering
	\begin{overpic}[width=\textwidth]{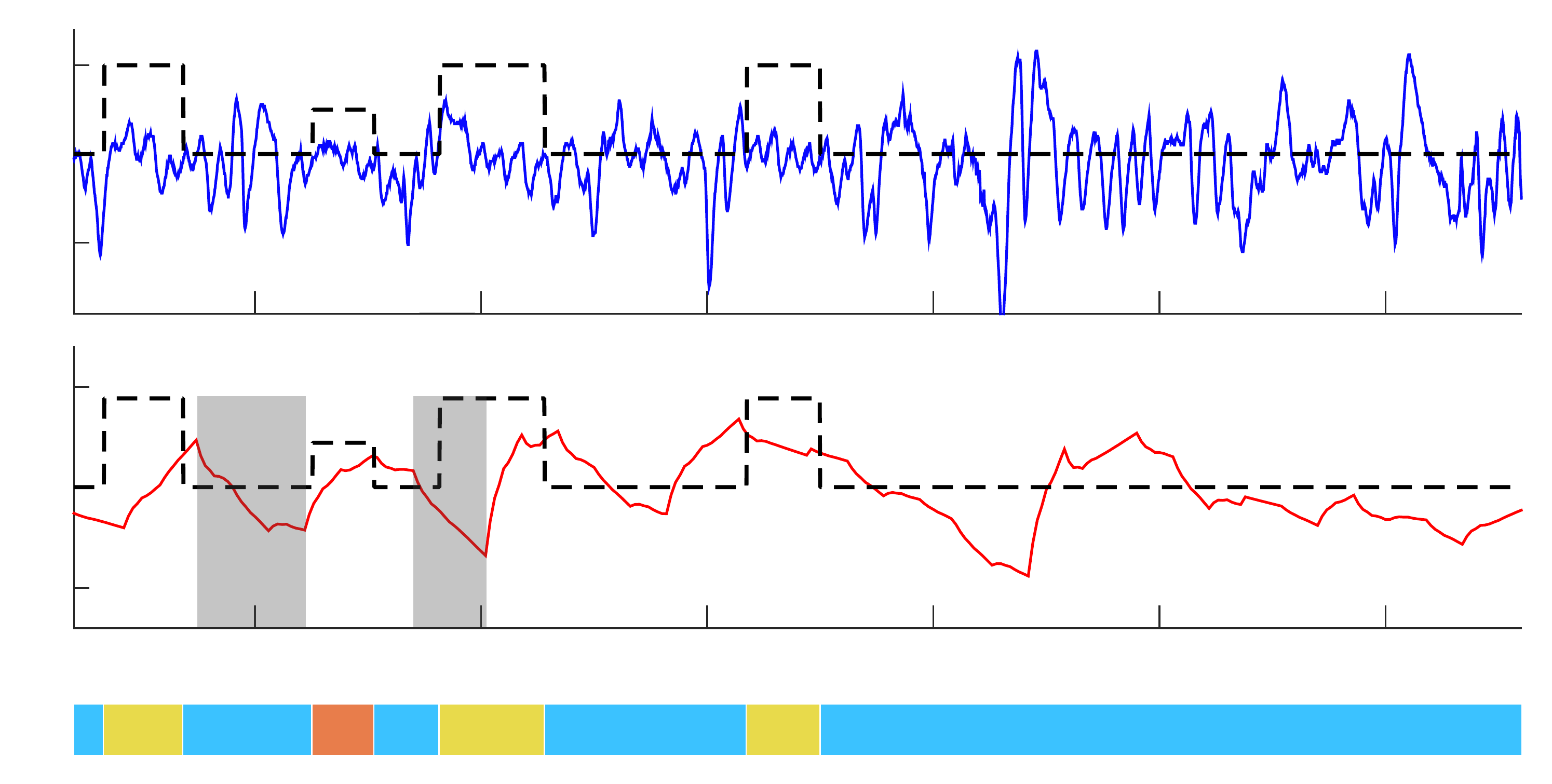}
		\put(4.9,2.8){\scriptsize N}
		\put(8.3,2.8){\scriptsize H}
		\put(15.3,2.8){\scriptsize N}
		\put(21.3,2.8){\scriptsize A}
		\put(25.4,2.8){\scriptsize N}
		\put(30.9,2.8){\scriptsize H}
		\put(41,2.8){\scriptsize N}
		\put(49.8,2.8){\scriptsize H}
		\put(74,2.8){\scriptsize N}
		\put(14,8){\scriptsize 13350}
		\put(28.4,8){\scriptsize 13400}
		\put(43,8){\scriptsize 13450}
		\put(57.7,8){\scriptsize 13500}
		\put(72.2,8){\scriptsize 13550}
		\put(86.8,8){\scriptsize 13600}
		\put(47,6){\scriptsize Time (sec.)}
		\put(2,45.7){\scriptsize 50}
		\put(3,40){\scriptsize 0}
		\put(1.4,34.3){\scriptsize -50}
		\put(3,25){\scriptsize 5}
		\put(3,18.5){\scriptsize 0}
		\put(2.4,12){\scriptsize -5}
		\put(-1,35){\scriptsize \rotatebox{90}{Airflow (raw)}}
		\put(-1,12.7){\scriptsize \rotatebox{90}{$\textrm{SaO}_2$ (filtered)}}
		\put(-1,0.6){\scriptsize \rotatebox{90}{Events}}
	\end{overpic}
	\caption{A small portion of an airflow signal (top), a wavelet filtered $\textrm{SaO}_2$ signal (middle) and labels of normal breathing and abnormal respiratory events (apnea and hypopnea) that occur during sleeping (bottom). Data obtained from \cite{quan_sleep_1997}.}
	\label{fig: oximetria_y_eventos_N_A_H}
\end{figure*}

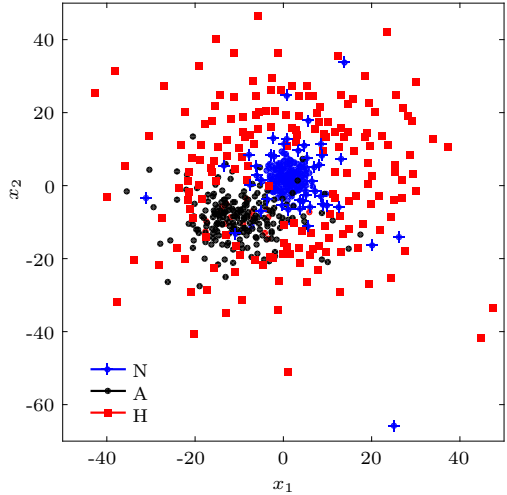
\begin{figure}[tbp]
	\centering
	\begin{overpic}[width=6.7cm]{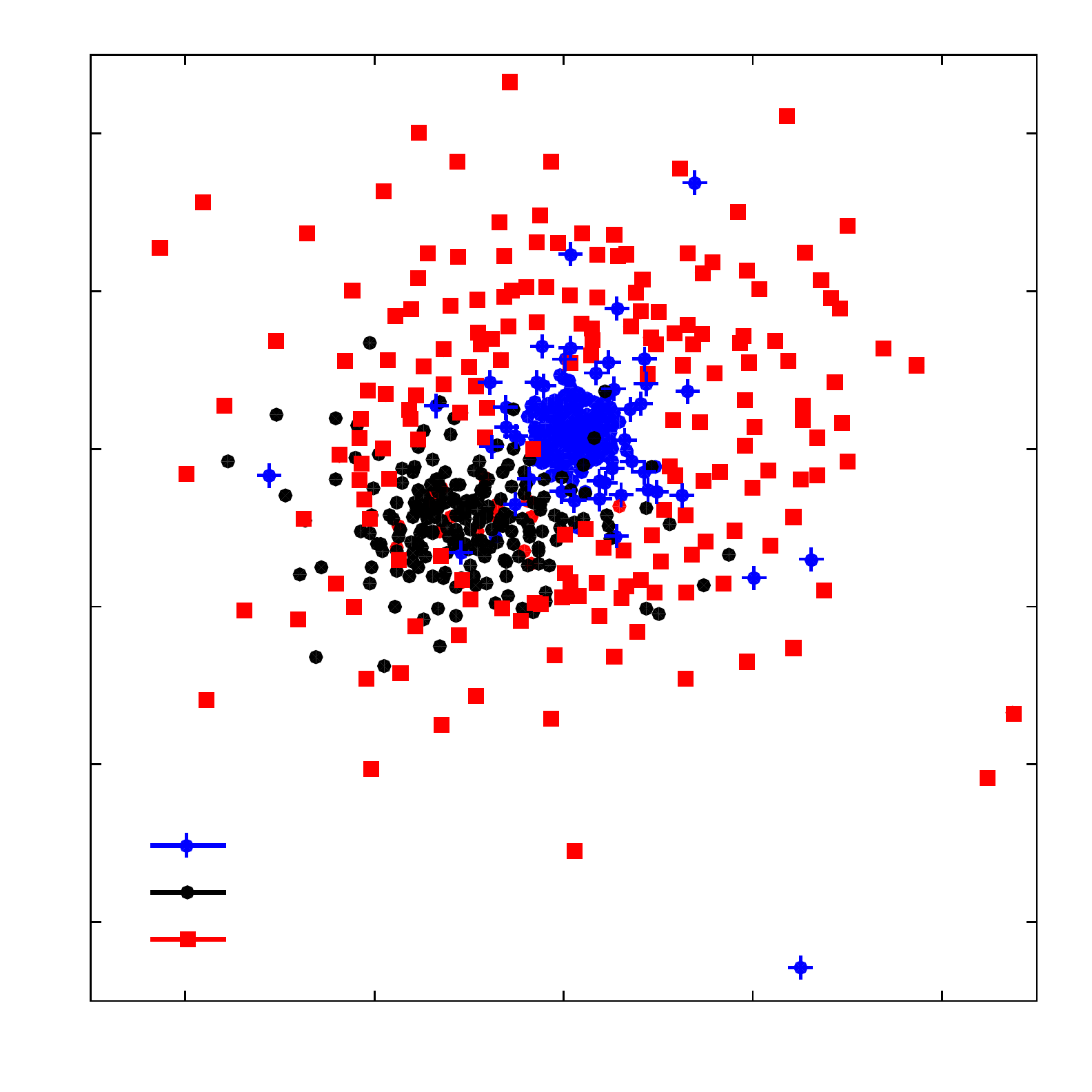}
		\put(13.3,3.8){\scriptsize -40}
		\put(30.7,3.8){\scriptsize -20}
		\put(50.6,3.8){\scriptsize 0}
		\put(66.8,3.8){\scriptsize 20}
		\put(84,3.8){\scriptsize 40}
		\put(1.1,14.2){\scriptsize -60}
		\put(1.1,28.5){\scriptsize -40}
		\put(1.1,43){\scriptsize -20}
		\put(4.7,57.4){\scriptsize 0}
		\put(2.5,72){\scriptsize 20}
		\put(2.5,86.6){\scriptsize 40}
		\put(49.5,-1.2){\scriptsize $x_1$}
		\put(-2,56){\scriptsize \rotatebox{90}{$x_2$}}
		\put(22,21){\scriptsize N}
		\put(22,16.5){\scriptsize A}
		\put(22,12){\scriptsize H}
	\end{overpic}
	\caption{A representation of the class distribution after applying a mapping denoted by \emph{Sammon mapping}, in its two most relevant attributes obtained from $\textrm{SaO}_2$ signals (estimated taking into account 200 examples for each class). Data obtained from \cite{quan_sleep_1997}.}
	\label{fig: dist_eventos_N_A_H}
\end{figure}

The severity of OSAH syndrome is classified as normal, mild, moderate or severe depending on whether the AHI values fall within the intervals $\left[0,5\right)$, $\left[5,15\right)$, $\left[15,30\right)$, or $\left[30,\infty \right)$, respectively. It is known that towards the end of each apnea or hypopnea event, a desaturation of the hemoglobin occurs. It is therefore reasonable to think that these deasaturations contain valuable information related the particular events of apnea and hypopnea, which are very often impossible to be recognized and distinguished by the human eye. The top and middle waveforms in Figure \ref{fig: oximetria_y_eventos_N_A_H} show a six-minutes portion of a typical airflow signal and the corresponding filtered $\textrm{SaO}_2$ signal, respectively (see Section \ref{sec: database and signal preprocesing}) \cite{rolon_discriminative_2017}. The labels N (normal breathing), A (apnea) and H (hypopnea) are shown at the bottom. It is important to mention that these labels were introduced by medical experts, after a detailed analysis of all the signals acquired during the PSG study. By observing both the airflow and the $\textrm{SaO}_2$ signals, it can be seen that the time frame between the reduction (or stopping) of airflow and the beginning of oxygen desaturation levels is very variable. The $\textrm{SaO}_2$ signal at the middle of Figure \ref{fig: oximetria_y_eventos_N_A_H} shows two gray-highlighted portions on the left, corresponding to the time intervals where desaturations produced by a hypopnea event (left) and an apnea event (right) occur.
As it can be observed, the minimum saturation values and the general morphology of the signal on those two intervals are very similar. Hence, it becomes evident that automatic recognition of single apnea and hypopnea events from only $\textrm{SaO}_2$ signals is a very challenging classification problem. To further visualize the difficulty of this classification problem, a technique for dimensionality reduction called ``Sammon Mapping'' was applied to low-dimensional samples of $\textrm{SaO}_2$ signals \cite{van2009dimensionality}. Figure \ref{fig: dist_eventos_N_A_H} shows projections to two-dimensional attributes of signals for the classes N, H and A. It can be observed that the distribution of the different classes in the attributes space highly overlap each other. Although the distributions representing both classes normal breathing and apnea events seems to be fairly separated, the distribution of hypopnea events presents a very high dispersion leading to a great degree of overlap with them.

\section{Dictionary Learning for Sparse Representation}\label{sec: dict learning for sr}
\subsection{Basic methods}\label{sec:non_discr_dict}
The representation of signals based on a dictionary consists of finding appropriate linear combinations of atoms in the prescribed dictionary to represent a given set of signals. This representation problem can be divided in two sub-problems: an inference problem and a learning problem. We proceed to describe each one of them. For that, let $\vec{x}\in\mathbb{R}^N$ be an input signal and let $\Phi\in\mathbb{R}^{N\times M}$ (usually $M \geq N$) be a dictionary whose columns $\vec{\phi}_j\in\mathbb{R}^N$, $j=1,2,\cdots,M$, are atoms that we want to use for representing $\vec{x}$ in the form $\vec{x} \cong \Phi \vec{a}=\sum_{j=1}^M a_j \phi_j$. Here, and in the sequel, we shall refer to the vector $\vec{a}=[a_1 \; a_2 \; \cdots \; a_M]^T \in\mathbb{R}^M$ as a ``representation'' of $\vec{x}$.

The inference problem essentially consists of finding the optimal (in a certain sense) representation $\vec{a}$ of the given signal $\vec{x}$. A sparse solution of this problem is a representation $\vec{a}$ with just a few non-zero components. If in a given representation a certain coefficient is non-zero, then we shall refer to it as an ``active'' component.

A way of obtaining a sparse representation of the signal $\vec{x}$ based on the dictionary $\Phi$ consists of solving the following problem: 
\begin{equation*}
\left(P_0\right) \;\;\; \vec{a}^* \doteq \underset{\vec{a}\in \mathbb{R}^M}{\textrm{argmin}} \; {||\vec{a}||}_{0}, \quad \textrm{subject to} \; \vec{x} = \Phi \vec{a},
\label{p_zero}
\end{equation*}
\noindent where $||\vec{a}||_0$ denotes the $l_0$ pseudo-norm, defined as the number of non-zero elements of $\vec{a}$.

Solving $\left(P_0\right)$ is generally an NP hard problem yielding this approach highly unsuitable for most applications \cite[\S 1.8]{elad_sparse_2010}. This is so because in $\left(P_0\right)$ we are imposing an exact representation which, in most practical cases, is neither strictly necessary nor desired. To overcome the computational burden which entails solving problem $\left(P_0\right)$, several relaxed versions of it have been considered. One of them consists of allowing a small representation error while imposing an upper bound on the $l_0$ pseudo-norm, i.e. solve:
\begin{equation*}
\left(P_0^q \right) \;\;\; \vec{a}^* \doteq \underset{\vec{a}\in \mathbb{R}^M}{\textrm{argmin}} \; ||\vec{x} - \Phi \vec{a}||_2, \quad \textrm{subject to} \; ||\vec{a}||_0 \leq q,
\label{p_zero_q}
\end{equation*}
\noindent where $q$ is a prescribed integer parameter. Several approaches for solving problem $\left(P_0^q\right)$ were proposed \cite{chen2001atomic,mallat_matching_1993,tropp_signal_2007}. The one most widely used is Orthogonal Matching Pursuit (OMP) which consists of approximating the solution in a greedy way providing a good trade-off between computational cost and representation error \cite{sahoo2015signal}. Additionally, the method ensures convergence to the projection of $\vec{x}$ into the span of the dictionary atoms \cite{tropp_signal_2007}.

The dictionary $\Phi$ can be constructed either using a pre-specified group of atoms (such as those obtained through the Wavelet Packet decomposition) or by means of data-driven learning approaches. The dictionary learning problem associated to the data $q$, $M$, $N \in \mathbb{N}$, $M \geq N$ and a collection of $n$ signals in $\mathbb{R}^N$, $\vec{x}_1,\cdots,\vec{x}_n$, can be formally written as:
\begin{equation*}
\left(\text{DL} \right) \;\;\; \left[\Phi^*,\vec{a}_1^*,\cdots, \vec{a}_n^* \right] \doteq \underset{\underset{\vec{a}_i \in \mathbb{R}^M,||\vec{a}_i||_0\leq q,1\leq i \leq n.}{\Phi \in \mathbb{R}^{N \times M}}}{\textrm{argmin}} \; \sum_{i=1}^n||\vec{x}_i-\Phi \vec{a}_i||_2^2
\label{p_dict_learning}
\end{equation*}
\noindent A solution of this problem yields on one hand a dictionary $\Phi$ and, on the other hand, representations $\vec{a}_i$ for all the signals $\vec{x}_1,\cdots,\vec{x}_n$ (in terms of such a dictionary) complying with the imposed sparsity constraint. Although several methods for solving (DL) exist, the most widely used is an iterative algorithm called K Singular Value Decomposition (KSVD) \cite{aharon_ksvd_2006}. This approach consists of two steps: an inference step and a dictionary learning step. The OMP algorithm (for example) is used for obtaining the representation coefficients, which is then followed by a dictionary learning step where the atoms are updated one-at-a-time and the representation coefficients are adjusted in order to minimize the total representation error.
\subsection{Discriminant dictionaries}\label{sec: discr_dict}
As mention above, a dictionary $\Phi$ can be constructed using data-driven learning methods aimed exclusively to minimize the total representation error. However, a dictionary learned in this way quite often produces representations of signals which turn out to be unsatisfactory if the final objective is pattern recognition. This is so because, as it is well known, a good representation does not necessarily guarantee good classification performance. A way to overcome this flaw consists of incorporating available prior information about class membership of the signals into the objective function in (DL) \cite{zhang_discriminative_2010,pham_joint_2008}. In \cite{zhang_discriminative_2010}, for example, a discriminant version of the standard KSVD method applied to face recognition was presented. In that work, the authors included a discriminant term into the objective function of the standard KSVD algorithm. Results have shown that such a modification constitutes an appropriate way to learn dictionaries simultaneously complying with both desired properties: low reconstruction error and high recognition rates. In \cite{pham_joint_2008}, a sparse-constrained optimization problem combining the objective function of the classification and the representation error of both labeled and unlabeled data, was formulated.

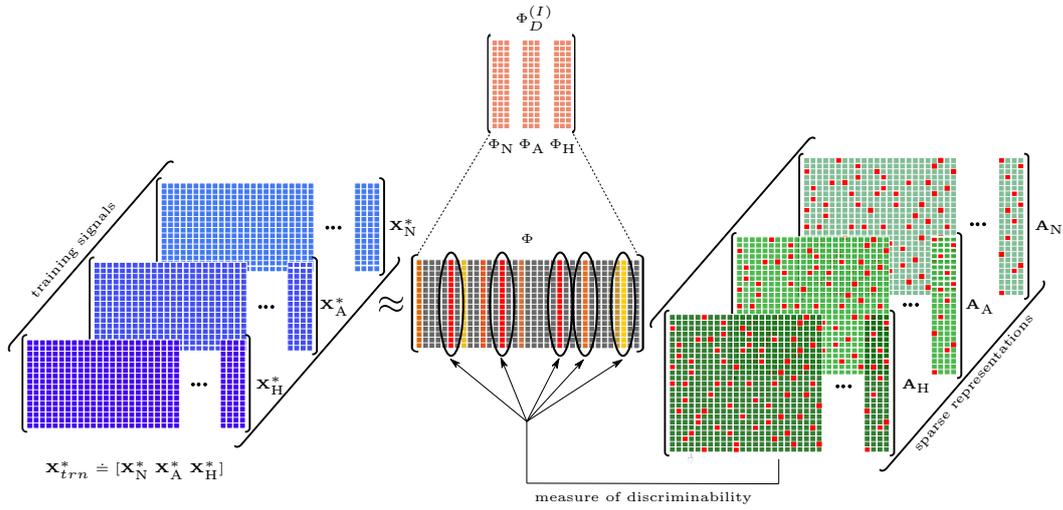
\begin{figure*}[tbp]
	\centering
	\begin{overpic}[width=14cm]{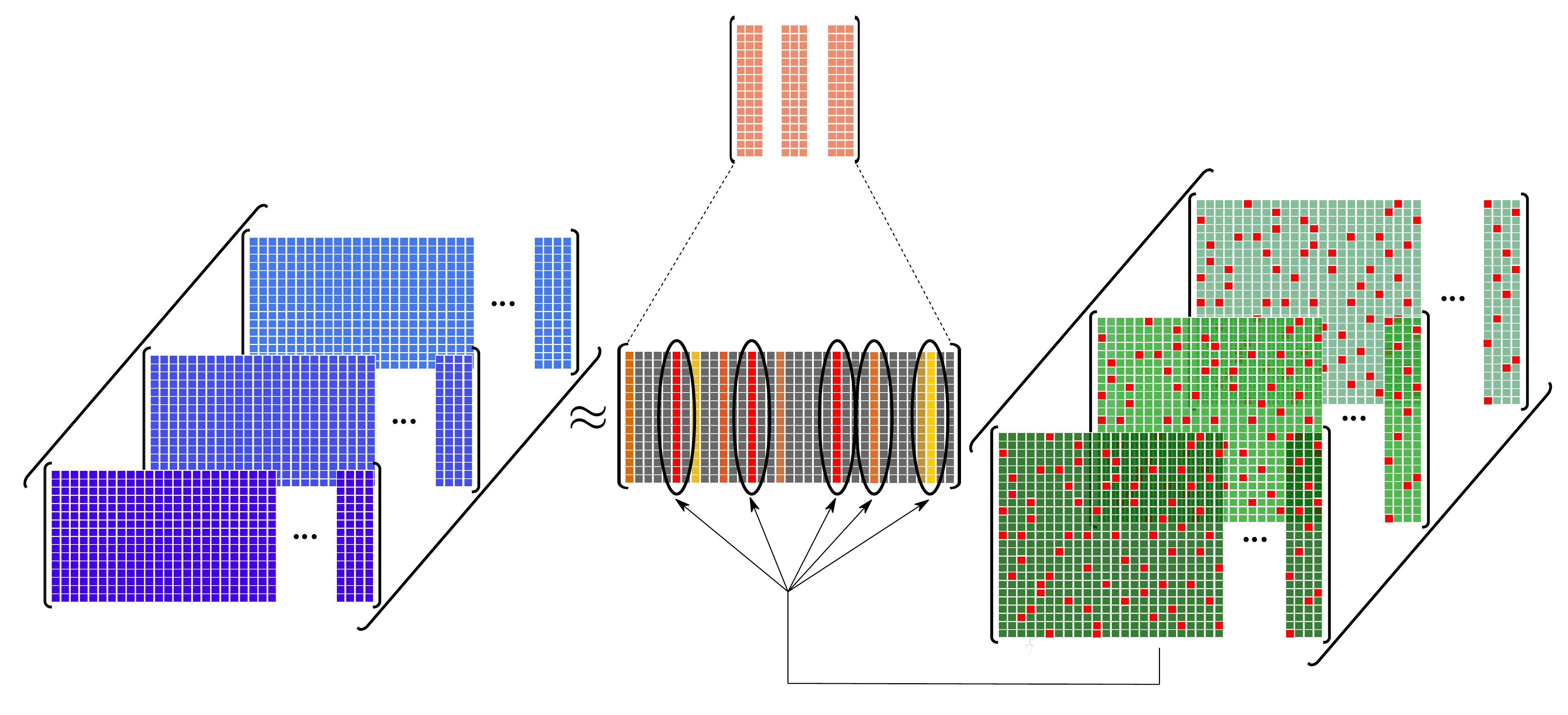}
		\put(3.5,19.5){\tiny \rotatebox{49.5}{training signals}}
		\put(49.7,24.8){\tiny $\Phi$}
		\put(49.2,45.6){\tiny $\Phi_D^{(I)}$}
		\put(5,3){\tiny $\vec{X}^*_{trn} \doteq [ \vec{X}^*_\textrm{N} \; \vec{X}^*_\textrm{A} \; \vec{X}^*_\textrm{H}]$}
		\put(86.5,5){\tiny \rotatebox{49.5}{sparse representations}}
		\put(37.3,26){\tiny $\vec{X}^*_\textrm{N}$}
		\put(30.8,18.5){\tiny $\vec{X}^*_\textrm{A}$}
		\put(24.8,11){\tiny $\vec{X}^*_\textrm{H}$}
		\put(46.6,33.7){\tiny $\Phi_\textrm{N}$}
		\put(49.5,33.7){\tiny $\Phi_\textrm{A}$}
		\put(52.4,33.7){\tiny $\Phi_\textrm{H}$}
		\put(51,0.3){\tiny measure of discriminability}
		\put(98,26){\tiny $\vec{A}_\textrm{N}$}
		\put(91.5,18.5){\tiny $\vec{A}_\textrm{A}$}
		\put(85.5,11){\tiny $\vec{A}_\textrm{H}$}
	\end{overpic}
	\caption{A schematic representation of the learning process of discriminant structured dictionaries using the MDCS method.}
	\label{block_diagram_3}
\end{figure*}

With the objective of improving classification performance, new approaches based on the design of structured dictionaries were recently proposed \cite{jiang_label_2013,rao_clustering_2012,chen_learn_2016,ataee2019structured}. A structured dictionary can be thought of as a collection of class-specific sub-dictionaries which are designed to capture discriminant properties of each class as well as common features among all classes in the data. In this direction, an initial approach consists of learning one dictionary for each class, then classify by minimizing the representation error among all classes \cite{wright_robust_2009}. Recently, a method called ``Most Discriminative Columns Selection'' (MDCS), which was shown to be capable of efficiently building structured dictionaries in a binary classification scheme, was developed \cite{rolon_discriminative_2017}. Figure \ref{block_diagram_3} shows a schematic representation of the MDCS procedure for a three-class classification problem. In this case the classes are identified as N, A and H. The dictionary $\Phi$ is learned in an unsupervised way using all training signals for solving problem (DL). After that, the representation matrices $\vec{A}_\textrm{N}$, $\vec{A}_\textrm{A}$ and $\vec{A}_\textrm{H}$ whose columns are the corresponding representation vectors, are computed using the three separate sets of labeled signals $\vec{X}^*_\textrm{N}$, $\vec{X}^*_\textrm{A}$ and $\vec{X}^*_\textrm{H}$, respectively. Next, the atoms of $\Phi$ are ranked according to a prescribed measure of discriminability in terms of their role in the sparse representation of the signals for each class (see Section \ref{sec: discr_criteria}). Following this ranking procedure, and given a prescribed positive integer $I$ (more on this later), the best $I$ atoms for each class are selected and used for building new class-specific sub-dictionaries $\Phi_\textrm{N}$, $\Phi_\textrm{A}$ and $\Phi_\textrm{H}$ for classes N, A and H, respectively. The structured dictionary, which we denote by $\Phi_D^{(I)}$, is finally constructed by stacking side-by-side all sub-dictionaries, i.e. $\Phi_D^{(I)}=[\Phi_\textrm{N} \; \Phi_\textrm{A} \; \Phi_\textrm{H}]$. The parameter $I$ is used to restrict the size of the final dictionary, in the sense that $\Phi_D^{(I)}$ will end up having exactly $I \times k$ columns, where $k$ is the number of classes. This restriction intends to improve the generalization capabilities reducing the size of the final feature vectors, what in turn, reduces the computing time required for classification.

Along MDCS, a method for discriminant features selection called ``Most Discriminative Atoms Selection'' (MDAS) was proposed \cite{rolon_discriminative_2017}. The main difference between both MDCS and MDAS is that in the later no new structured dictionary $\Phi_D^{(I)}$ is built. Instead the original dictionary $\Phi$ is preserved and the ranking of the atoms is used only to select the components to be used for classification. It is important to point out that although both MDCS and MDAS were originally proposed for dealing only with binary classification problems, their extension to multi-class problems is straight forward. In what follows, we shall denote by MDCS-BC, MDCS-MC, MDAS-BC and MDAS-MC the binary and multiclass versions of MDCS and MDAS, respectively.

Following on the idea behind MDCS, an iterative extension of it naturally emerges. In this sense, a new method called ``Discriminant Atom Selection KSVD'' (DAS-KSVD) was recently proposed \cite{rolon_a_multiclass_2018}. This method is suitable for multi-class classification problems and it can be thought of as a generalization of MDCS. The main difference with MDCS is that, instead of selecting all $I$ class-specific atoms in a unique step, DAS-KSVD chooses only one discriminant atom for each one of the classes at each step. Additionally, DAS-KSVD incorporates a re-sampling technique which promotes diversity in the generation of the discriminant atoms. This re-sampling process requires of a prescribed parameter $\tau_1$, $0\leq \tau_1<1$, for adjusting the sampling probability of all training signals. It is important to mention that all sampled signals are degraded by incorporating additive noise of magnitude proportional to $l\tau_2$, where $l$ is the number of iterations and $\tau_2 \in \left[0,1\right)$ is another prescribed parameter. For more details about these re-sampling and signal degradation procedures, we refer the reader to \cite{rolon_a_multiclass_2018}. The steps for constructing the dictionary with DAS-KSVD are summarized in Algorithm \ref{dksvd_algorithm} below.
\begin{algorithm}[tbp]
	\caption{DAS-KSVD method}\label{dksvd_algorithm}
	\begin{algorithmic}[1]
		\Procedure{das-ksvd}{$\vec{X}_{trn},q,r_f,I,\vec{c},t$}
		\State $p_0(i)=1/n$, for all $i$
		\For{$l \gets 0, I-1$} 
		\State $\left[ \vec{X}_{lrn},p_{l+1} \right] \gets $~\Call{SampleData}{$\vec{X}_{trn},t,p_l,l$}
		\State $\Phi \gets $~\Call{Ksvd}{$\vec{X}_{lrn},r_f,q$}
		\State $\vec{A}_{lrn} \gets $~\Call{OMP}{$\vec{X}_{lrn},\Phi,q$}
		\State $m_{\alpha^*,\beta^*} \gets $~\Call{DiscMeasure}{$\vec{A}_{lrn},\vec{c},q$}
		\State $\Phi_d \gets $~\Call{GetAtoms}{$\Phi,m_{\alpha^*,\beta^*}$}
		\State $\Phi_D^{(i)} \gets $~\Call{SaveAtoms}{$\Phi_d$}
		\EndFor
		\State \textbf{return} $\Phi_D^{(I)}$
		\EndProcedure
	\end{algorithmic}
\end{algorithm}

\begin{figure*}[htbp]
	\centering
	\begin{overpic}[width=14cm]{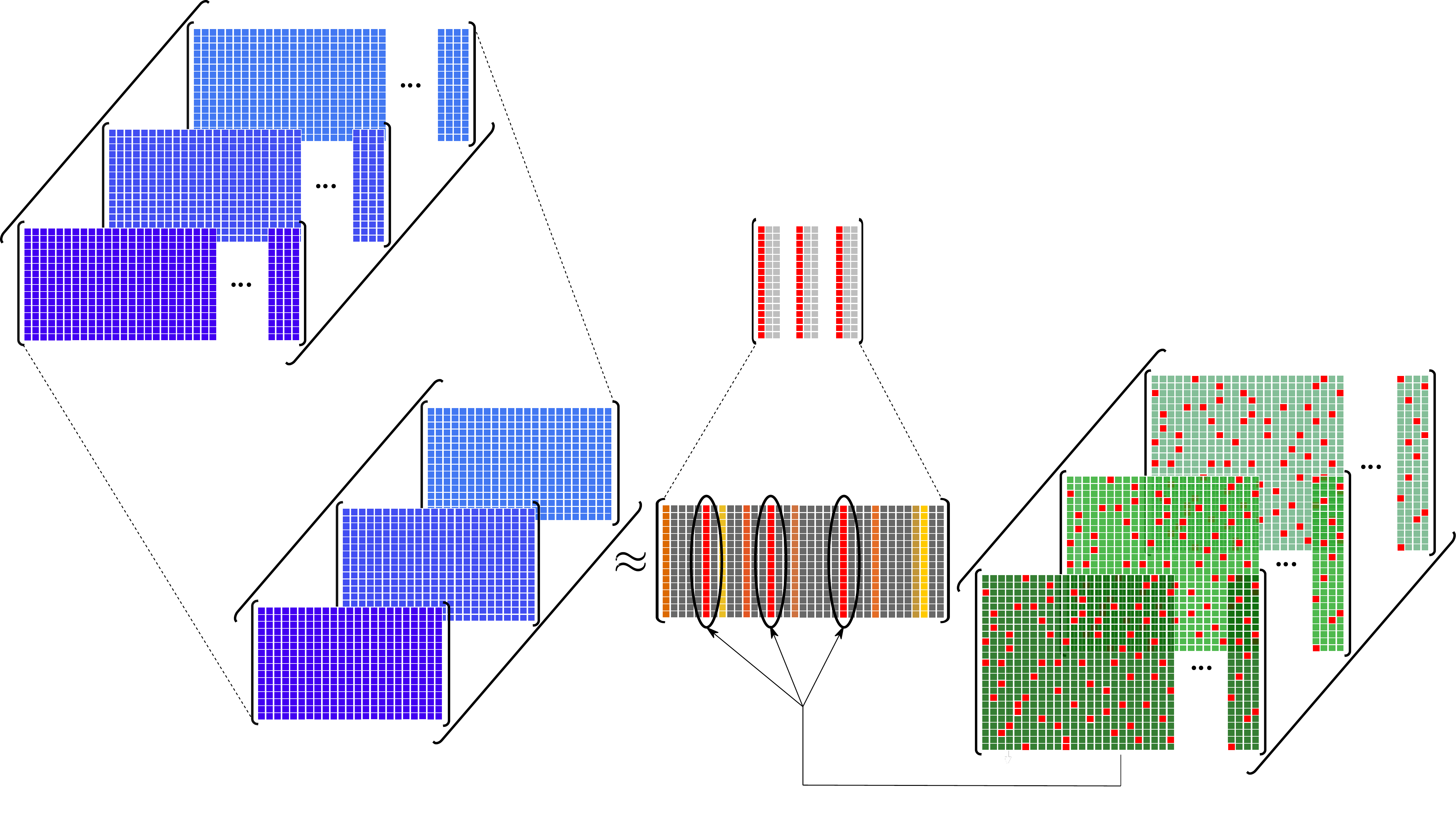}
		\put(2,45){\tiny \rotatebox{49.5}{training signals}}
		\put(17.8,19){\tiny \rotatebox{49.5}{learning signals}}
		\put(54.5,23.5){\tiny $\Phi$}
		\put(7.3,25){\tiny \rotatebox{-59}{re-sampling process}}
		\put(53.8,43.2){\tiny $\Phi_D^{(I)}$}
		\put(88,5){\tiny \rotatebox{49.5}{sparse representations}}
		\put(4.5,29.5){\tiny $\vec{X}_{trn} \doteq [ \vec{X}_\textrm{N} \; \vec{X}_\textrm{A} \; \vec{X}_\textrm{H}]$}
		\put(15.8,3){\tiny $\vec{X}_{lrn} \doteq [ \hat{\vec{X}}_\textrm{N} \; \hat{\vec{X}}_\textrm{A} \; \hat{\vec{X}}_\textrm{H}]$}
		\put(43,25){\tiny $\hat{\vec{X}}_\textrm{N}$}
		\put(37.5,18){\tiny $\hat{\vec{X}}_\textrm{A}$}
		\put(31.4,11){\tiny $\hat{\vec{X}}_\textrm{H}$}
		\put(51.6,31.7){\tiny $\Phi_\textrm{N}$}
		\put(54.3,31.7){\tiny $\Phi_\textrm{A}$}
		\put(57,31.7){\tiny $\Phi_\textrm{H}$}
		\put(55.3,1){\tiny measure of discriminability}
		\put(99,25){\tiny $\vec{A}_\textrm{N}$}
		\put(93.2,18.2){\tiny $\vec{A}_\textrm{A}$}
		\put(87.4,11){\tiny $\vec{A}_\textrm{H}$}
		\put(33,50.9){\tiny $\vec{X}_\textrm{N}$}
		\put(27.1,43.7){\tiny $\vec{X}_\textrm{A}$}
		\put(21.4,37){\tiny $\vec{X}_\textrm{H}$}
	\end{overpic}
	\caption{A schematic representation of one iteration of the learning process of discriminant structured dictionaries using the DAS-KSVD method.}
	\label{block_diagram_4}
\end{figure*}
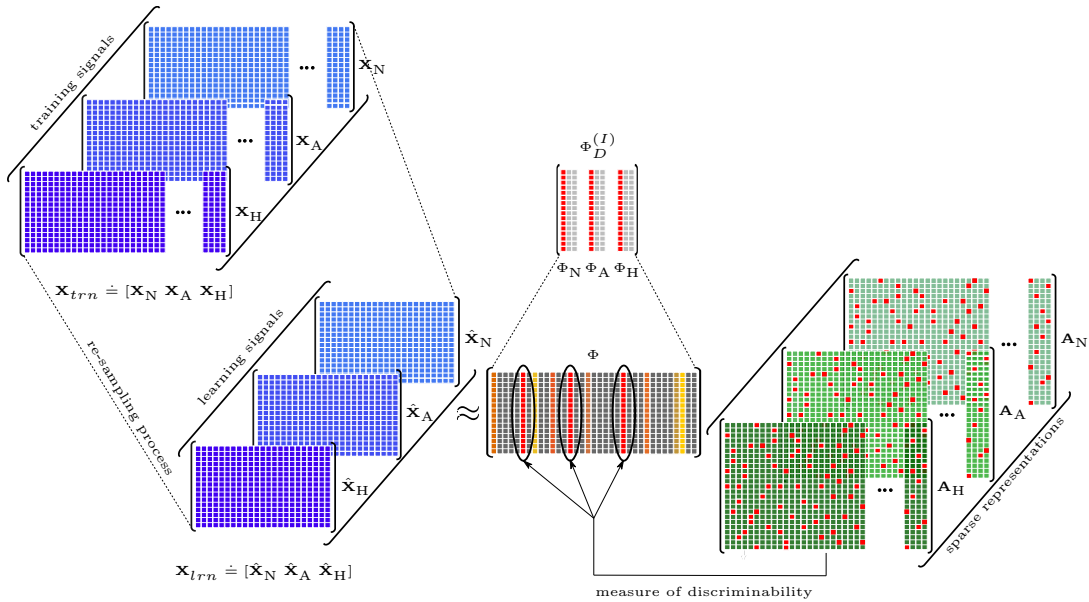

Figure \ref{block_diagram_4} shows a schematic representation of one iteration of DAS-KSVD for a three-class classification problem. Observe that before using a method for solving (DL), a re-sampling technique is applied. Then, the dictionary $\Phi$ is learned in an unsupervised way using all learning signals $\vec{X}_{lrn}$.
After that, the representation matrices $\vec{A}_\textrm{N}$, $\vec{A}_\textrm{A}$ and $\vec{A}_\textrm{H}$ whose columns are the corresponding representation vectors, are computed using the three separate sets of learning signals $\hat{\vec{X}}_\textrm{N}$, $\hat{\vec{X}}_\textrm{A}$ and $\hat{\vec{X}}_\textrm{H}$, respectively. Next, the atoms of $\Phi$ are ranked according to an appropriately defined multi-class measure of discriminability (details about this measure are presented in Section \ref{sec: discr_criteria}). After this ranking procedure, only one atom for each class is selected and used for building new class-specific sub-dictionaries $\Phi_\textrm{N}$, $\Phi_\textrm{A}$ and $\Phi_\textrm{H}$ for classes N, A and H, respectively. The structured dictionary, which is denoted by $\Phi_D^{(I)}$, is finally constructed by stacking side-by-side all sub-dictionaries, i.e. $\Phi_D^{(I)}=[\Phi_\textrm{N} \; \Phi_\textrm{A} \; \Phi_\textrm{H}]$.
\subsection{Discriminant criteria}\label{sec: discr_criteria}
As previously mentioned, both MDCS and DAS-KSVD require of measures for quantifying the discriminant capabilities of each one of the dictionary atoms. The detection of atoms containing useful discriminant information can be addressed in different ways. Among all existing alternatives, the most commonly used strategy consists of performing comparisons between conditional probability distributions \cite{lin1991divergence,rolon2018complexity}. In what follows, we proceed to describe two different criteria that shall be used in this article.

Based on the idea that the discriminant atoms of a dictionary are those more frequently used for representing signals belonging to a particular class, a measure called ``Discriminative Conditional Activation Frequency'' (DCAF) was proposed \cite{rolon_discriminative_2017,rolon2018complexity}. This measure was shown to be capable of efficiently quantifying the discriminability of the atoms in the context of a binary classification problem. The approach essentially consists of using the conditional activation probability $p_\ell^j$ of the atom $\phi_j$ given the class $\ell$. This conditional activation probability, which is defined as $p_\ell^j \doteq P(a_j \not = 0|\vec{x} \in \ell)$, can be approximated efficiently by the quotient $\eta_\ell^j/n_\ell$, where $\eta_\ell^j$ and $n_\ell$ are the conditional activation frequency (number of times that the atom $\phi_j$ becomes active for representing class $\ell$ signals) and the number of class $\ell$ signals, respectively. To quantify the discriminant capability of an atom $\phi_j$, the absolute value of the difference of its conditional activation probabilities for classes $\ell=1$ and $\ell=2$ $(|p_1^j-p_2^j|)$ is computed. This value will be close to one if (an only if) the atom $\phi_j$ becomes much more active for one class only and, in that case, it can be thought of as a quantifier of the capability of $\phi_j$ to provide important information for signal classification. In addition, observe that DCAF is symmetric, its value is always non-negative and is inexpensive in terms of computing time. Finally, if the classes are balanced, DCAF can be computed just by counting the number of times that each atom becomes active without dividing by the number of class-specific signals.

With the objective of extending DCAF to a more general framework, a new strategy to detect discriminant atoms in the context of multi-class classification problems was recently proposed \cite{rolon_a_multiclass_2018}. The approach consists of defining and using a new multi-objective function $m_{\alpha,\beta}$ aimed at quantifying the discriminant properties of each one of the atoms in a given dictionary. This function is defined as a convex combination of three discriminant terms, all based on the affine sparse representations of the data. In what follows, we proceed to describe each one of such terms.

For a given $j$, $1 \leq j \leq M$, we denote by $\ell_j^+$ the class that maximizes all conditional activation probabilities $p_\ell^j$, for all $\ell=1,2,\cdots,k$. If there is more than one value of $\ell$ maximizing $p_\ell^j$, $\ell_j^+$ is defined by randomly choosing one of them, for instance the smallest one (note that the order of the classes is completely irrelevant). Similarly, for a fixed $j$, $1 \leq j \leq M$, $\ell_j^*$ is defined as the class leading to the second largest conditional activation probability. Here again if there is more than one value of $\ell_j^*$ satisfying that condition, then $\ell_j^*$ is randomly chosen among them.

Next, the function $m_{af}: \; \{1,2,\cdots,M\} \rightarrow \mathbb{R}_0^+$, known as the ``activation frequency'' measure, is defined by
\begin{equation}
m_{af}(j) \doteq \frac{p_{\ell_j^+}^j-p_{\ell_j^*}^j}{p_{\ell_j^+}^j}.
\end{equation}
Note that $0 \leq m_{af}(\cdot) \leq 1$. The atom $\phi_j$ is said to be discriminant (for class $\ell_j^+$) if and only if $m_{af}(j)>0$. Clearly, within this setting, if an atom $\phi_j$ is discriminant, it will be so only for the class $\ell_j^+$. Moreover, the value of $m_{af}(j)$ can be thought of as a ``measure'' of the degree of discriminability of the atom $\phi_j$ (for the corresponding class $\ell_j^+$), based solely on the conditional activation frequency information.

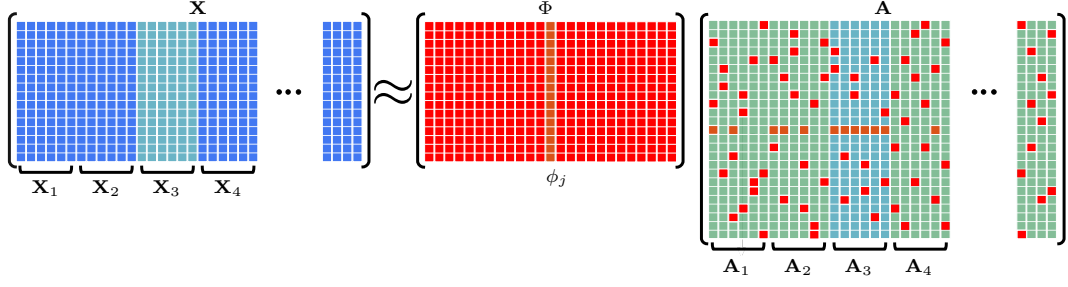
\begin{figure*}[htbp]
	\centering
	\begin{overpic}[width=15cm]{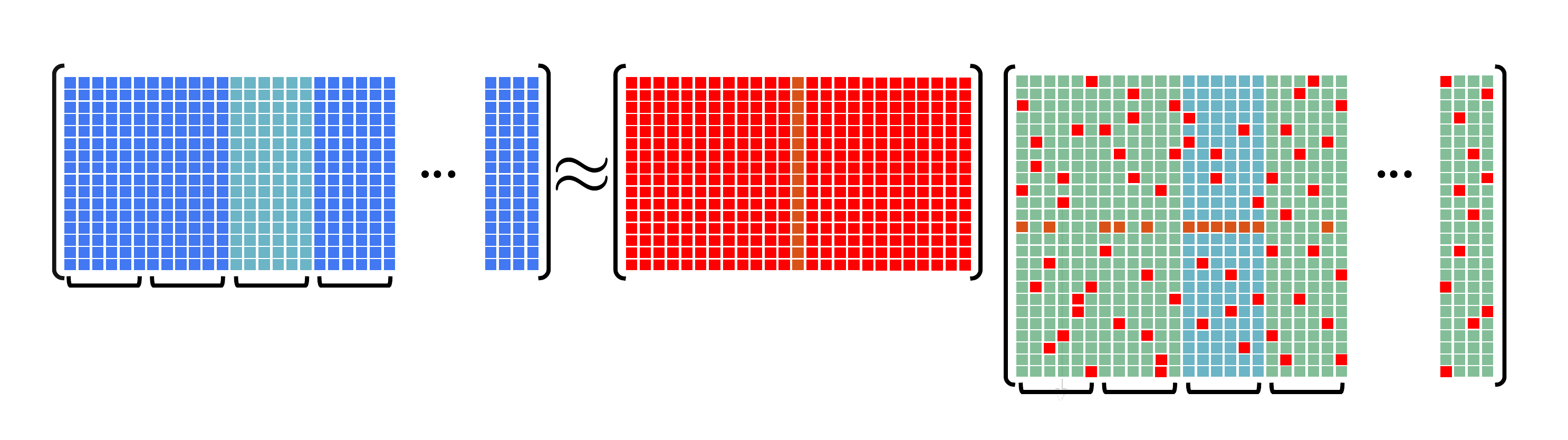}
		\put(5.4,8){\scriptsize $\vec{X}_1$}
		\put(10.7,8){\scriptsize $\vec{X}_2$}
		\put(16,8){\scriptsize $\vec{X}_3$}
		\put(21.4,8){\scriptsize $\vec{X}_4$}
		\put(66,1){\scriptsize $\vec{A}_1$}
		\put(71.4,1){\scriptsize $\vec{A}_2$}
		\put(76.7,1){\scriptsize $\vec{A}_3$}
		\put(82,1){\scriptsize $\vec{A}_4$}
		\put(50.5,9){\scriptsize $\phi_j$}
		\put(19.2,23.8){\scriptsize $\vec{X}$}
		\put(49.8,23.8){\scriptsize $\Phi$}
		\put(79.3,23.8){\scriptsize $\vec{A}$}
	\end{overpic}
	\caption{An illustration of the atoms activations by taking into account signals coming from 4 different classes.}
	\label{fig: ilustracion_act_disc_atom}
\end{figure*}

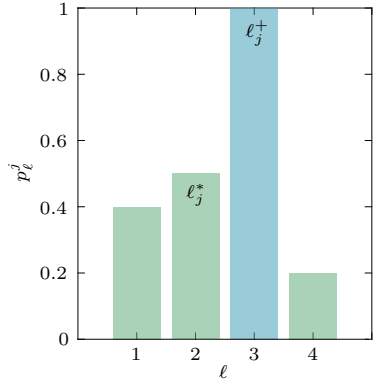
\begin{figure}[htbp]
	\centering
	\begin{overpic}[width=5cm]{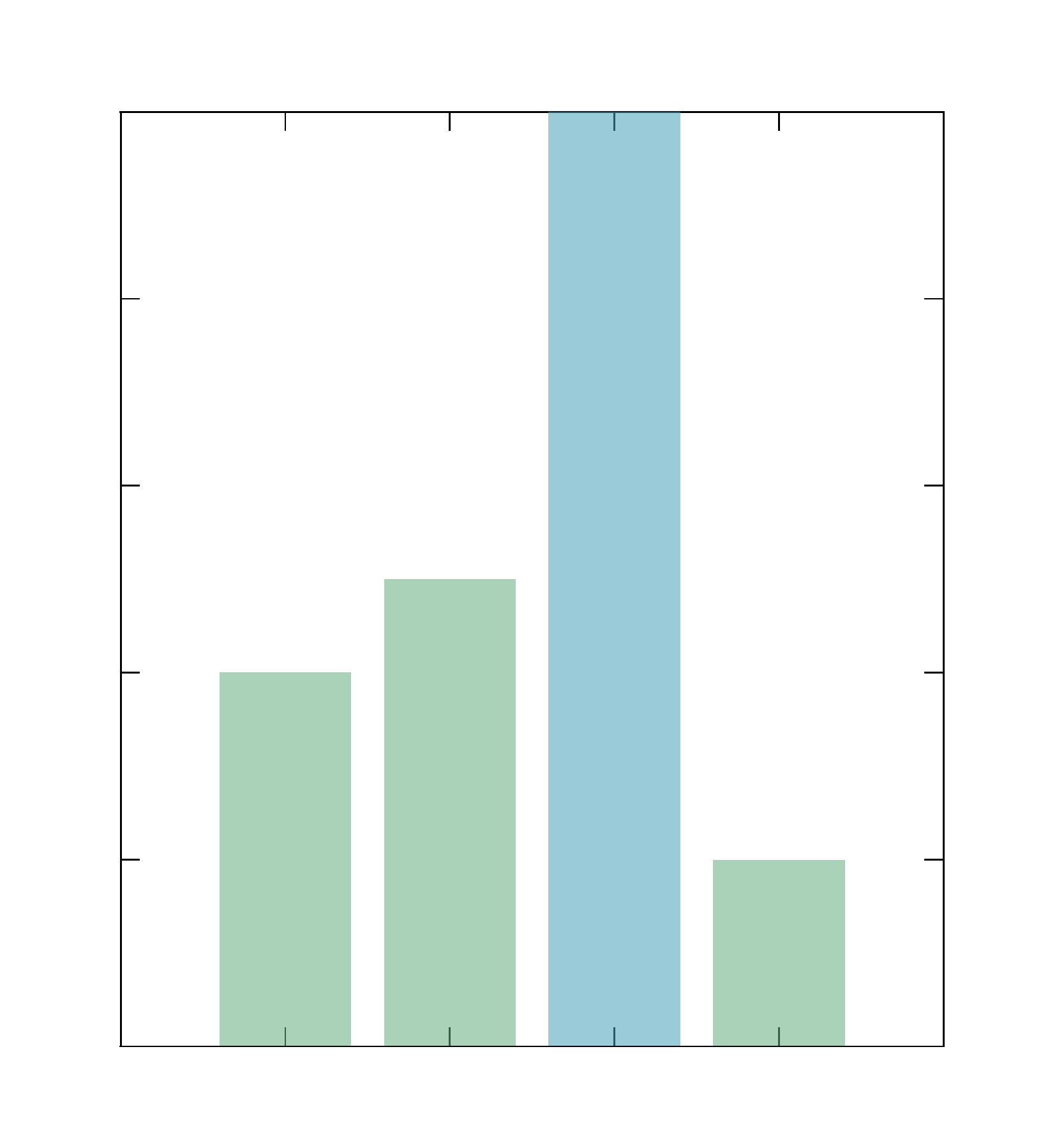}
		\put(23.5,3.5){\scriptsize 1}
		\put(38,3.5){\scriptsize 2}
		\put(52.3,3.5){\scriptsize 3}
		\put(66.7,3.5){\scriptsize 4}
		\put(5.8,88.6){\scriptsize 1}
		\put(1.2,72.3){\scriptsize 0.8}
		\put(1.2,56){\scriptsize 0.6}
		\put(1.2,39.6){\scriptsize 0.4}
		\put(1.2,23.5){\scriptsize 0.2}
		\put(5.8,7.2){\scriptsize 0}
		\put(45,-1){\scriptsize $\ell$}
		\put(-6.5,48.5){\scriptsize \rotatebox[]{90}{$p_\ell^j$}}
		\put(51.5,83){\scriptsize $\ell_j^+$}
		\put(37,43.5){\scriptsize $\ell_j^*$}
	\end{overpic}
	\caption{A schematic representation of conditional activation probabilities of a certain atom $\phi_j$ for each one of the classes.}
	\label{fig: barras_frec_act_cond}
\end{figure}

The conditional activation probability of the atoms can be graphically illustrated by taking into account sparse representations of signals coming from different classes in terms of a given dictionary. For that, we shall consider a signal matrix $\vec{X}$, which comprises four different class-labeled signals, i.e. $\vec{X} \doteq \left[ \vec{X}_1 \; \vec{X}_2 \; \vec{X}_3 \; \vec{X}_4 \right]$. Moreover, let $\vec{A} \doteq \left[ \vec{A}_1 \; \vec{A}_2 \; \vec{A}_3 \; \vec{A}_4 \right]$ be the matrix which provides a sparse representation of $\vec{X}$ in terms of the dictionary $\Phi$, through $\vec{X} \approx \Phi \vec{A}$. Figure \ref{fig: ilustracion_act_disc_atom} shows a graphic representation of sparse representations $\lbrace \vec{A}_\ell \rbrace$, for $\ell=1,2,3,4$, of all signal matrices $\lbrace \vec{X}_\ell \rbrace$ in terms of a dictionary $\Phi$. In particular, both $\phi_j$ and its corresponding active components $a_j$ (the active elements that are part of the $j^{\textrm{th}}$-row of $\vec{A}$) have been highlighted in orange.
It can be observed that $\phi_j$ becomes more frequently active for signals of class $\ell=3$ than for all the others. Also, since $\phi_j$ is always used to represent class $\ell=3$ signals, the conditional activation probability of $\phi_j$ given the class $\ell=3$ is maximum, i.e. $p_3^j=1$ and therefore $\ell_j^+=3$. On the other hand, it is easy to determine that $\ell=2$ is the class leading to the second largest conditional activation probability and, in this case, $p_3^j=0.5$ and therefore $\ell_j^*=2$. Hence, according to the activation frequency measure, $m_{af}(j)=0.5$. Additionally, Figure \ref{fig: barras_frec_act_cond} shows a bar plot representing each one of the conditional activation probability values of the atom $\phi_j$.

Besides providing useful information regarding the activation of the atoms, the sparse representation of signals is capable of efficiently highlighting intrinsic properties and relevant class-related features of the data. With this observation in mind, a second criterion that takes into account the magnitude of the representation coefficients is presented. For that, given an atom $\phi_j$, let $\ell_j^+$ and $\ell_j^*$ be as before, and let $\vec{A}_{\ell_j^+}$ and $\vec{A}_{\ell_j^*}$ be matrices providing sparse representations of $\vec{X}_{\ell_j^+}$ and $\vec{X}_{\ell_j^*}$, respectively, in terms of the dictionary $\Phi$, i.e. $\vec{X}_{\ell_j^+}=\Phi \vec{A}_{\ell_j^+}$ and $\vec{X}_{\ell_j^*}=\Phi \vec{A}_{\ell_j^*}$. Additionally, let $q_\ell^j$ denote the quotient $||\left[\vec{A}_\ell\right]_{j,:}||_1/n_\ell$, where $\left[\vec{A}_\ell\right]_{j,:}$ represents the $j^\textrm{th}$-row of the matrix $\vec{A}_\ell$. Then, the ``coefficient magnitude'' measure is the function $m_{cm}: \; \{1,2,\cdots,M\} \rightarrow \mathbb{R}_0^+$ defined by
\begin{equation}
m_{cm}(j) \doteq \frac{q_{\ell_j^+}^j-q_{\ell_j^*}^j}{q_{\ell_j^+}^j}.
\end{equation}
Here again $0 \leq m_{cm}(\cdot) \leq 1$. Based on this measure, an atom $\phi_j$ is said to be discriminant (for the class $\ell_j^+$) if and only if $m_{cm}(j)>0$ and, in that case, the value of $m_{cm}(j)$ quantifies the ability of $\phi_j$ to discriminate class $\ell_j^+$ data, according to this criterion.

We now proceed to describe a third criterion for quantifying the discriminant degree of the atoms. This criterion takes into account the contribution of each atom $\phi_j$, $j=1,2,\cdots,M$, to the total representation error. Let $\vec{A}_\ell \doteq [\vec{a}_1 \; \vec{a}_2 \; \cdots \vec{a}_{n_\ell}]$ be a matrix providing a sparse representation of $\vec{X}_\ell \doteq [\vec{x}_1 \; \vec{x}_2 \; \cdots \vec{x}_{n_\ell}]$. The total representation error for all class $\ell$ signals when $\phi_j$ is removed can then be written as $||\vec{E}_\ell^j||_F^2 = ||\vec{X}_\ell-\sum_{i \not= j}\phi_i \left[\vec{A}_\ell\right]_{i,:}||_F^2$ (for more details we refer the reader to \cite{aharon_ksvd_2006}). A large value of $||\vec{E}_\ell^j||_F^2$ indicates that $\phi_j$ is a highly discriminant atom. Then, the ``representation error'' measure $m_{re}: \; \{1,2,\cdots,M\} \rightarrow \mathbb{R}_0^+$ is defined by
\begin{equation}
m_{re}(j) \doteq \frac{r_{\ell_j^*}^j-r_{\ell_j^+}^j}{r_{\ell_j^*}^j},
\end{equation}
\noindent where $r_\ell^j \doteq ||\vec{E}_\ell^j||_F^2 / n_\ell$, for $\ell=1,2,\cdots,k$, $j=1,2,\cdots,M$. It is clear that $0 \leq m_{re}(\cdot) \leq 1$, and an atom $\phi_j$ is said to be discriminant (for class $\ell_j^+$) with respect to this criterion if and only if $m_{re}(j)>0$.

Each one of the three previously mentioned criteria quantifies the discriminant properties of an atom from three different perspectives. It is then reasonable to think of a criterion that properly combines all three of them. With that in mind, given two positive parameters $\alpha$ and $\beta$, with $\alpha+\beta \leq 1$, the combined discriminant measure $m_{\alpha,\beta}: \; \{1,2,\cdots,M\} \rightarrow \mathbb{R}_0^+$ is defined as
\begin{eqnarray}\label{function_m}
m_{\alpha,\beta}(j) \doteq \alpha \, m_{af}(j)+ \beta \, m_{cm}(j)+ (1-\alpha-\beta) \, m_{re}(j).
\end{eqnarray}
Clearly, as $\alpha$ and $\beta$ vary between $0$ and $1$, (\ref{function_m}) exhausts all possible convex combinations of the three single measures $m_{af}$, $m_{cm}$ and $m_{re}$. A challenging problem that immediately arises is to find the ``optimal'' pair of parameters $(\alpha^*,\beta^*)$ leading to the best recognition rate. However, up to our knowledge, no analytical method exist for finding $(\alpha^*,\beta^*)$. For this reason, in this article a discrete search for such a pair of parameters in the $\alpha-\beta$ plane is performed. 
\section{Experimental setup}\label{sec: exp setup}
The main objective of this article is the comparison of the overall classification performances in the context of OSAH syndrome screening of MDCS, MDAS (both in their binary and multiclass versions) and DAS-KSVD. To achieve that objective, two experiments were carried out. The first one was designed with the final goal of classifying the segments of $\textrm{SaO}_2$ signals in one and only one of the three classes: normal breathing (N), apnea (A) or hypopnea (H). The second experiment was designed to detect the existence or non-existence of the pathology. The whole experimental setup is described below.
\subsection{Database and signal pre-processing} \label{sec: database and signal preprocesing}
The Sleep Heart Health Study (SHHS) database was originally designed to explore possible correlations between sleep related breathing disorders and cardiovascular diseases \cite{quan_sleep_1997,lind_recruitment_2003}. This database consists of several complete PSG studies, each one of them containing a group of physiological signals such as EEG, ECG, nasal airflow and $\textrm{SaO}_2$. In addition, annotations of sleep stages, arousals and events of apnea and hypopnea are provided. The criteria that medical experts adopted for identifying apnea and hypopnea events were the following \cite{berry2012rules}. An apnea event is a complete (or almost complete) blockage of the upper airflow for at least ten seconds, usually associated with a desaturation in the $\textrm{SaO}_2$ signal or an arousal. A hypopnea event is a reduction in airflow by less than a 70\% of the baseline level, associated with a desaturation in the $\textrm{SaO}_2$ signal or an arousal.

In this article we make use of the first online version of the database called ``Sleep Heart Health Study'' (SHHS-2)\footnote{\url{https://physionet.org/physiobank/}}. This database contains 995 complete PSG studies, 41 of which were discarded due to labeling flaws \cite{rolon_discriminative_2017}. Among the remaining 954 studies, a set of 667 (70\%) were randomly selected for training purposes. The remaining 287 (30\%) were left out for the final test.

Mainly due to patient movements, baseline wander and undesired disconnections (among many other factors), the original raw $\textrm{SaO}_2$ signals require of an appropriate pre-conditioning process. For that, linear interpolation and wavelet filters, as those used in a previous work \cite{rolon_discriminative_2017}, were applied. Figure \ref{fig: oximetria_y_eventos_N_A_H} shows a small portion of a $\textrm{SaO}_2$ signal (top) and its wavelet-filtered version (middle).

Signals are segmented into vectors $\vec{x}_i \in \mathbb{R}^{N}$ of length $N=128$ (corresponding to 128 seconds of the signal recording) with a 75\% overlapping between two consecutive segments. In this process, segments containing artifacts or disconnections are discarded. Then, a matrix $\vec{X}_{trn} \in \mathbb{R}^{128 \times n_{trn}}$ is constructed by stacking side-by-side $n_N$, $n_A$ and $n_H$ vectors belonging to the classes N, A and H, respectively. Clearly, $n_{trn}=n_N+n_A+n_H$. Similarly, another matrix $\vec{X}_{tst} \in \mathbb{R}^{128 \times n_{tst}}$ is built using the vectors associated to the testing set.
\subsection{Dictionary learning settings}\label{sec: struct dict learning}
For DAS-KSVD, all experiments were performed setting $I=20$ (i.e. 20 iterations). Thus, the final structured dictionary consists of 60 atoms (assuming $k=3$). For each one of the classes used to learn the full dictionary (by means of KSVD), the number of samples was set to $t=500$. Also, several trials were performed in order to obtain adequate values for both parameters $\tau_1$ and $\tau_2$. In particular, it was found that values of $\tau_1=0.5$ and $\tau_2=0.1$ are suitable for this application. In addition, $\tau_2=0.1$ resulted in the best trade-off between signal degradation and the number of iterations. Finally the optimal pair of parameters $(\alpha^*,\beta^*)$ was found to be $(0.33,0.17)$. All parameters of the KSVD method such as the sparsity constrain $q$ and the redundancy factor of the dictionary $r_f$, were set equal to those used in a previous work \cite{rolon_a_multiclass_2018}. Finally, for both MDCS-MC and MDAS-MC, all parameters were set as for DAS-KSVD. It is important to mention, however, that these two methods make use of a different input data matrix $\vec{X}_{trn}^*$  which is composed of a balanced set of randomly selected segments from $\vec{X}_{trn}$. Since $n_L$ segments were chosen for each class, the final size of $\vec{X}_{trn}^*$ was $128\times 3n_L$ where $n_L$ is the number of segments chosen from each class.
\subsection{Classification}\label{sec: neural network}
In order to classify segments of $\textrm{SaO}_2$ signals into the three different classes, a feed-forward Multilayer Perceptron (MLP) neural network was used. In particular the experiments were run using three layers (input, hidden and output). Naturally, input and output layer sizes were set to 60 and 3 corresponding to $I \times k$ and $k$, respectively. The hidden layer consisted of 500 neurons with a \emph{tansig} activation function. To train this network, conjugate gradient descent was used. For classification purposes, the cost function was chosen as the mean squared error (MSE).

To carry out the first experiment, two balanced sets of 21000 and 4500 samples were randomly selected from $\vec{X}_{trn}$ and used for training and validation purposes, respectively. Also, an additional balanced set of 4500 samples was randomly chosen from $\vec{X}_{tst}$ and used for testing purposes. Then, sparse representations of these new datasets in terms of the previously learned dictionary were found and used as input of the classifier.
\subsection{Detection of OSAH syndrome}
In a typical PSG study, the recorded signals are provided to medical experts who identify and label apnea and hypopnea events, which are later used for computing the AHI index. In a similar way, in our analysis, each testing study was appropriately filtered and segmented in order to classify its segments as N, A and H, by means of the previously described process. Then, an estimated AHI ($\textrm{AHI}_{\textrm{est}}$) was computed by counting the total number of segments classified as A or H and dividing it by the duration of the study, in hours. This new index was used for OSAH syndrome detection. Finally, each study was considered as pathological if the obtained $\textrm{AHI}_{\textrm{est}}$ was greater than a certain prescribed detection threshold \cite{rolon2018complexity}.
\subsection{Performance measures}\label{sec: performance measures}
To analyze and quantify the ability of the MLP to classify segments of $\textrm{SaO}_2$ signals in a multiclass scenario, a confusion matrix was constructed. The confusion matrix is a very useful tool for reporting results in multiclass classification problems because it gives a full overview of all relations between the classifier predictions and the known (true) labels. Rows and columns of such a matrix refer to known and predicted class labels of the dataset, respectively, while its diagonal and off-diagonal elements correspond to observations that are correctly and incorrectly classified, respectively. This information summarizes the types of errors that occur during training, validation and testing. Based on the confusion matrix, the overall accuracy as well as other three widely used class-specific measures (sensitivity (Se), specificity (Sp) and precision (Pr)) were extracted. In this article, the confusion matrix is normalized by dividing each one of the elements in its rows by the total number of testing samples that belong to each class.

To assess the ability of the proposed system in detecting patients suspected of suffering from moderate to severe OSAH syndrome, i.e. persons having an AHI index greater than 15, a Receiver Operating Characteristics (ROC) analysis was performed \cite{swets_roc_1979}. The optimal cut-off point (associated to a prescribed detection threshold) of the ROC curve is the one that simultaneously maximizes sensitivity and specificity. Also, the accuracy (Acc) and the area under the ROC curve (AUC) were computed.
\section{Results and discussions}\label{sec: results and discussions}
This section presents a qualitative description of the atoms learned by DAS-KSVD as well as the findings achieved through the experiments described above: classification of segments and detection of OSAH syndrome. 
\subsection{A qualitative analysis}
\begin{figure}[htbp]
	\centering
	\begin{overpic}[width=6.5cm]{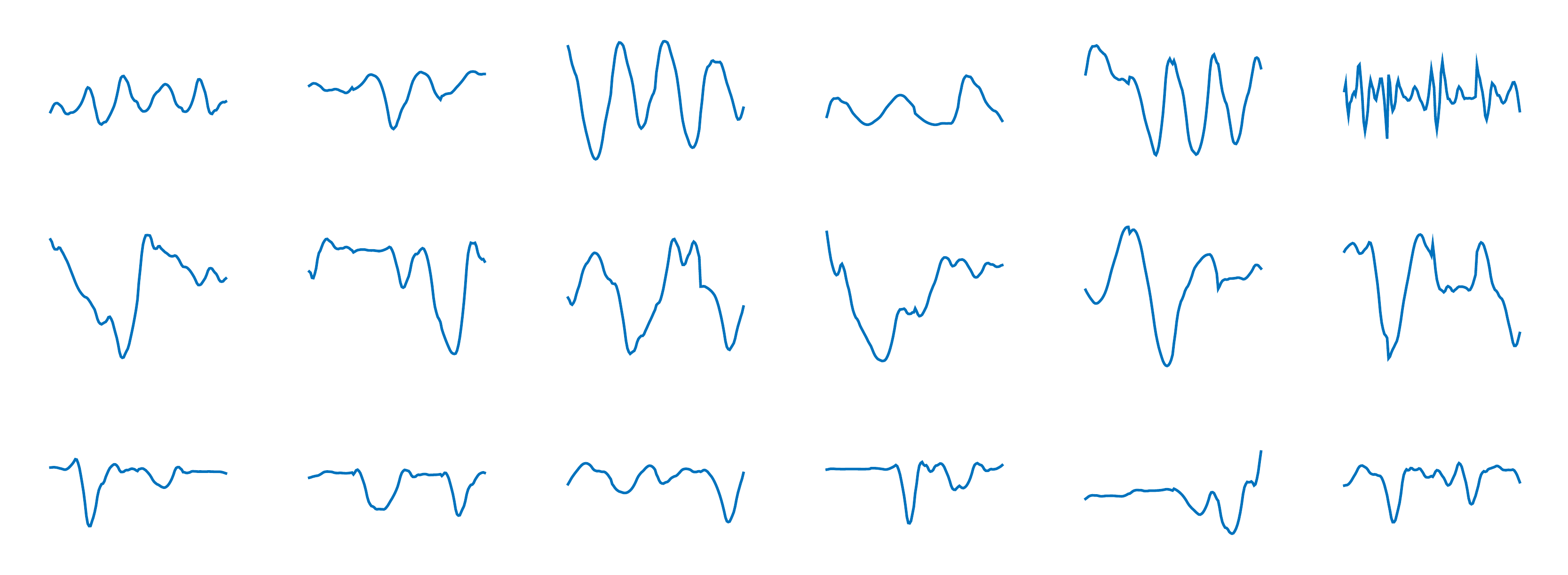}
		\put(-5,29.5){\scriptsize $\Phi_\textrm{N}$}
		\put(-5,18){\scriptsize $\Phi_\textrm{A}$}
		\put(-5,5){\scriptsize $\Phi_\textrm{H}$} 
	\end{overpic}
	\caption{Typical atoms corresponding to $\Phi_\textrm{N}$ (top), $\Phi_\textrm{A}$ (middle) and $\Phi_\textrm{H}$ (bottom).}
	\label{fig: disc_atoms_N_A_H}
\end{figure}

DAS-KSVD was used to learn a structured dictionary for $\textrm{SaO}_2$ signals using the procedures and parameters described in Section \ref{sec: exp setup}. A structured dictionary $\Phi_D^{(I)}=\left[\Phi_\textrm{N} \; \Phi_\textrm{A} \; \Phi_\textrm{H}\right]$ of size $128 \times 60$ was obtained. Figure \ref{fig: disc_atoms_N_A_H} shows the waveforms of some representative atoms corresponding to each one of the discriminant dictionaries $\Phi_\textrm{N}$ (upper), $\Phi_\textrm{A}$ (middle) and $\Phi_\textrm{H}$ (bottom). 
A detailed analysis shows that each one of the dictionaries is built with atoms that capture particular types of class-related information. For instance, as it can be seen, most atoms in $\Phi_\textrm{N}$ present quite regular waveforms associated to inhalation-exhalation related changes in the oxygen saturation. On the other hand, atoms of $\Phi_\textrm{A}$, representing apnea events, present noticeable low frequency fluctuations. In pulse oximetry this is a typical behavior associated to the absence of respiratory airflow for a relatively long period of time. Finally, atoms of $\Phi_\textrm{H}$, associated to hypopnea events, reflect abnormal breathing through irregular patterns in pulse oximetry. It is timely to mention that all atoms of the dictionary are normalized so that their $\ell_1$-norm is equal to unity.
\subsection{Classification} \label{sec: det_of_indiv_events}
Features generated by DAS-KSVD were used to assess the ability of the MLP in classifying segments of $\textrm{SaO}_2$ signals. Table \ref{tab: multiclass_performance_matrix_measures} shows the confusion matrix constructed using all testing samples (left) and a summary of all class-specific performance measures extracted from such a matrix (right). The elements in the diagonal of Table \ref{tab: multiclass_performance_matrix_measures} (left) represent the normalized true positive rates. As it can be seen, the algorithm achieved true positive rates of 86.09\%, 63.20\% and 23.36\% for the classes N, A and H, respectively, resulting in an overall accuracy of 57.55\%. Note that if we were to limit our analysis only to the classes N and A (i.e. without tacking into account the third row and the third column of the confusion matrix), then the inter-class confusions would be relatively small. From the analysis of all these results several remarks can be drawn. First, DAS-KSVD constitutes a reasonable approach for classifying normal (breathing) and apnea events in pulse oximetry. Second, the results fall short of being good for detecting hypopnea events. In fact, more than half of them are misclassified as belonging to class N and more than one fourth are misclassified as belonging to class A. This last remark, however, is consistent with the results obtained using the {\it Sammon mapping} (see Section \ref{sec: sleep apnea} and Figure \ref{fig: dist_eventos_N_A_H}) where we saw that the projections of class A and N segments into the first two most important attributes of the mapping are clearly well separated, while the projections of class H segments overlap the other two classes and present a wide variance. 

\begin{table}[h]
	\caption{Normalized multiclass confusion matrix obtained using DAS-KSVD for segment classification (left) and the corresponding performance measures (right).}
	\label{tab: multiclass_performance_matrix_measures}
	\renewcommand{\arraystretch}{1.3}
	\centering
	\newcommand\items{3}   
	\noindent\begin{tabular}{cc*{\items}{|E}|}
		\arrayrulecolor{white} 
		\multicolumn{1}{c}{} &\multicolumn{1}{c}{}     &\multicolumn{\items}{c}{Predicted} \\ \hhline{~*\items{|-}|}
		\multicolumn{1}{c}{} & 
		\multicolumn{1}{c}{} & 
		\multicolumn{1}{c}{N} & 
		\multicolumn{1}{c}{A} & 
		\multicolumn{1}{c}{H} \\ \hhline{~*\items{|-}|}
		\multirow{\items}{*}{\rotatebox{90}{Known}} 
		&N  & 86.09   & 4.26  & 9.65   \\ \hhline{~*\items{|-}|}
		&A  & 21.33   & 63.20  & 15.46   \\ \hhline{~*\items{|-}|}
		&H  & 50.74   & 25.90   & 23.36   \\ \hhline{~*\items{|-}|}
	\end{tabular}
	\qquad	\qquad
	\setlength{\tabcolsep}{4pt}
	\begin{tabular}{*{4}{c}}
		\hline
		Class & Se (\%) & Sp (\%) & Pr (\%) \\
		\hline\hline
		N & 86.09 & 64.17 & 55.65\\
		A & 63.20 & 85.24 & 68.15\\
		H & 23.36 & 87.49 & 47.19\\
		\hline
	\end{tabular}
\end{table}



In order to gain insight into the reasons why DAS-KSVD outperforms all other evaluated approaches for OSAH syndrome detection (see next section), we compared its performance with that of MDCS-BC in classifying segments of $\textrm{SaO}_2$ signals as containing an event or not (i.e. without tacking into account whether it is an apnea or a hypopnea). It is important to mention that MDCS-BC was chosen because it achieved the best performance among all previously developed methods. In order to analyze the performance of DAS-KSVD in the binary classification problem, we unified labels of segments belonging to the classes A and H which led to a new (and unique) class denoted by A+H. Table \ref{tab: segment_detection_performance_measures} shows a summary of the performance of DAS-KSVD and MDCS-BC using all testing samples. It is important to point out that, in this case, the target class is A+H. As it can be observed, although both methods yielded similar sensibility percentages, DAS-KSVD reached a significantly better specificity and precision percentages than MDCS-BC. In other words, DAS-KSVD has become more specific having fewer false positives than the other one. This clearly indicates that in the classification process, segments that were misclassified as N, are now correctly classified as H.

\begin{table}[t]
	\renewcommand{\arraystretch}{1.3}
	\caption{Performance measures for A+H event detection from segments of $\textrm{SaO}_2$ signals using DAS-KSVD and MDCS-BC.}
	\label{tab: segment_detection_performance_measures}
	\centering
	\renewcommand{\arraystretch}{1.3}
	\setlength{\tabcolsep}{4pt}
	\begin{tabular}{*{4}{c}}
		\hline
		Method & Se (\%) & Sp (\%) & Pr (\%) \\
		\hline\hline
		DAS-KSVD & 64.42 & 86.09 & 89.83\\
		MDCS-BC & 64.15 & 78.99  & 71.23\\
		\hline
	\end{tabular}
\end{table}

\subsection{Detection of OSAH syndrome}
In this article, besides analyzing the ability of DAS-KSVD to classify segments of $\textrm{SaO}_2$ signals into the classes N, A and H, we make use of these predictions to detect the presence of the pathology (according to a prescribed AHI diagnostic threshold). In that sense, a comparison between DAS-KSVD with many other state-of-the-art methods in the diagnosis of moderate to severe OSAH syndrome ($\textrm{AHI}>15$) was performed. Table \ref{tab: detection of OSAH syndrome} shows a comparative summary of the results achieved by DAS-KSVD, MDCS-BC, MDCS-MC, MDAS-BC and MDAS-MC, and by the approaches introduced by Chiner \textit{et al.} \cite{chiner_nocturnal_1999}, V\'{a}zquez \textit{et al.} \cite{vazquez_automated_2000} and Schlotthauer \textit{et al.} \cite{schlotthauer_screening_2014}. It is important to point out that all results presented here were obtained using the same data partitions.
\begin{table}[htbp]
	\renewcommand{\arraystretch}{1.3}
	\caption{Performance measures for moderate to severe OSAH syndrome detection using different methods.}
	\label{tab: detection of OSAH syndrome}
	\centering
	\setlength{\tabcolsep}{4pt}
	\begin{tabular}{*{5}{c}}
		\hline
		Method & AUC & Se(\%) & Sp(\%) & Acc(\%)\\
		\hline\hline
		DAS-KSVD & \textbf{0.957} & \textbf{87.56} & \textbf{88.32} & \textbf{87.94}\\
		MDCS-MC & 0.942 & 86.13 & 86.62 & 86.37\\
		MDAS-MC & 0.913 & 82.32 & 82.78 & 82.55\\
		\hline
		MDCS-BC \cite{rolon_discriminative_2017} & 0.937 & 85.65 & 85.92 & 85.78\\
		MDAS-BC \cite{rolon_discriminative_2017} & 0.891 & 81.02 & 83.10 & 82.06\\
		Schlotthauer \textit{et al.} \cite{schlotthauer_screening_2014} & 0.922 & 84.11 & 85.94 & 85.02\\
		V\'{a}zquez \textit{et al.} \cite{vazquez_automated_2000} & 0.909 & 80.84 & 87.50 & 84.17\\
		Chiner \textit{et al.} \cite{chiner_nocturnal_1999} & 0.795 & 76.17 & 78.12 & 77.15\\
		\hline
	\end{tabular}
\end{table}

As can be observed in Table \ref{tab: detection of OSAH syndrome}, DAS-KSVD outperforms all other methods in both binary and multiclass versions. The application of DAS-KSVD resulted in an AUC value of 0.957 and a sensitivity, specificity and accuracy of 87.56\%, 88.32\% and 87.94\%, respectively. The method leading to the second largest performances is the multiclass version of MDCS (MDCS-MC). This method achieved an AUC value of 0.942 and a sensitivity, specificity and accuracy of 86.13\%, 86.62\% and 86.37\%, respectively. In addition, if we compare the results yielded by DAS-KSVD with those obtained by MDCS-MC, then it can be concluded that DAS-KSVD significantly enhances the overall performance achieved by MDCS-MC (assuming a $p$-value of 0.05).

It is important to note that, in most cases, multiclass-based methods outperform binary-based ones in the detection of the pathology. In fact, the application of MDCS-MC and MDAS-MC resulted in better performances than the ones yielded by their respective binary versions. For instance, MDAS-MC obtained an AUC value of 0.913 representing an improvement of 2.2\% regarding MDAS-BC, which achieved an AUC value of 0.891. Similarly, MDCS-MC yielded an improvement of 0.5\% with respect to MDCS-BC. On the other hand, it is appropriate to mention that although MDAS-MC shows improvements regarding MDAS-BC, its overall performance remains still below that of MDCS-BC and Schlotthauer \textit{et al}.

A more comprehensive analysis of Table \ref{tab: detection of OSAH syndrome} indicates that, although most discriminant methods achieved good results, DAS-KSVD outperforms all the others. The application of this method results in an area under the ROC curve of 0.957 as well as sensitivity and specificity of 87.56\% and 88.32\%, respectively. According to the original labels and taking into account a detection threshold of 15, the whole testing partition (287 studies) contains 216 and 71 studies diagnosed as pathological and normal (or healthy) patients, respectively. A 87.56\% sensitivity indicates that of the 216 pathological cases, 189 were correctly detected (true positive) while 27 were false positive. On the other hand, an 88.32\% specificity indicates that of the 71 healthy cases, 62 were appropriately identified (true negative) while only 9 were false negative. It is timely to note that for the 9 cases that DAS-KSVD yielded an AHI higher than 15, most events identified by the medical expert were hypopneas and most of them were not associated with noticeable desaturations in the $\textrm{SaO}_2$ signal. This fact indicates that the final scoring process was carried out following the AASM criteria. Hence, this issue may be one of the causes that led to the misclassification of hypopneas, since its distribution highly overlaps with the one corresponding to class N segments. Finally, if we look at the $\textrm{SaO}_2$ signal, there are a lot of cases where it becomes difficult to distinguish between class H and N segments.

\section{Conclusions}\label{sec: conclusions}
In this article, with the objective of OSAH syndrome screening, we applied a previously developed method called DAS-KSVD to classify segments of $\textrm{SaO}_2$ signals into normal breathing and abnormal respiratory events in a multiclass scenario. It was found that the combined discriminant measure, which is used by DAS-KSVD in the process of building the structured dictionary, is capable of efficiently selecting the most discriminant atoms for each one of the classes. In addition, DAS-KSVD yielded a structured dictionary composed by three sub-dictionaries each one associated to a particular class. We evaluated DAS-KSVD in two different but related applications, namely, classification of abnormal respiratory events and detection of moderate to severe OSAH syndrome. Although it is a very challenging task, the proposed method has demonstrated to be efficient for automatically discriminating between apnea and hypopnea events in a multiclass scheme. To detect the presence or absence of events, DAS-KSVD resulted more specific than the most competitive binary-based approach (MDCS-BC). This improvement is due to the ability of DAS-KSVD in separating between (apnea or hypopnea) events and normal breathing. In a similar way, the application of DAS-KSVD led to  the best reported performance in OSAH syndrome screening using a well known and publicly available database. This fact constitutes a strong evidence that our approach could be helpful in the development of new intelligent technologies for portable OSAH syndrome screening devices.

\section*{Acknowledgment}
The authors would like to acknowledge the financial support of Consejo Nacional de Investigaciones Cient\'{i}ficas y T\'{e}cnicas, CONICET, of the Air Force Office of Scientific Research, AFOSR /SOARD, through Grant FA9550-14-1-0130, of the Universidad Nacional del Litoral through projects CAI+D 50120110100519 and CAI+D 5012011010 0525 and of the Universidad Tecnol\'{o}gica Nacional through projects TEUTIPA0004711TC and ICUTIPA0007803TC.


\section{References}
\bibliographystyle{unsrt}

\end{document}